\documentclass[conference,a4paper]{APSIPA2020}
\usepackage{cite,caption,multirow,url}
\usepackage{graphicx}
\usepackage{amsmath}
\usepackage[psamsfonts]{amssymb}
\usepackage{amsxtra}
\usepackage{threeparttable}

\begin{document}

\title{A PRIVACY-PRESERVING CONTENT-BASED IMAGE RETRIEVAL SCHEME ALLOWING MIXED USE OF ENCRYPTED AND PLAIN IMAGES}

\author{%
\authorblockN{%
Kenta Iida\authorrefmark{1} and
Hitoshi Kiya\authorrefmark{1}
}
\authorblockA{%
\authorrefmark{1}
Tokyo Metropolitan University \\
6-6, Asahigaoka, Hino-shi, Tokyo, Japan\\
E-mail: iida-kenta2@ed.tmu.ac.jp, kiya@tmu.ac.jp}
}

\maketitle
\thispagestyle{empty}

\begin{abstract}
In this paper, we propose a novel content based-image retrieval scheme allowing the mixed use of encrypted and plain images for the first time.
In the proposed scheme, images are encrypted by a block-scrambling method developed for encryption-then-compression (EtC) systems.
The encrypted images, referred to as EtC images, can be compressed with JPEG, as well as for plain images.
Image descriptors used for the proposed retrieval is designed to avoid the effect of image encryption.
As a result, the use of EtC images and the descriptors allows us to carry out retrieval of both encrypted images and plain ones.
In an experiment, the proposed scheme is demonstrated to have the same performance as conventional retrieval methods with plain images, even under the mixed use of plain images and EtC ones.
\end{abstract}

\section{Introduction}
\label{sec:intro}
With the rapid growth of cloud computing, outsourcing images to cloud storage services and sharing photos have greatly increased.
Generally, images are uploaded and stored in a compressed form to reduce the amount of data.
In addition, most im†ages include sensitive information, such as personal data and copyright information \cite{Intro1, Intro2}.
However, cloud providers are not trusted in general, so there is the possibility of data leakage and unauthorized use in cloud environments.
Therefore, various privacy-preserving image identification, retrieval, and processing schemes have been studied for untrusted cloud environments \cite{enc1, EIR1,EIR4,EIR8,EIR5,EIR6,EIR2,EIR7,EIR9,EIR3,EIR10,EIR13,EIR11,EIR12,ESIMPLE, PPDNN1,PPDNN2,EID,add1}.

For the above reasons, privacy-preserving image-retrieval methods should satisfy generally three requirements: 1) protecting visual information on plain images, 2) having a high retrieval performance for encrypted images, and 3) being applicable to compressible encrypted images.
In regards to requirement 1), it is classified into two requirements: 1-a) protecting images stored in databases of cloud providers, and 1-b) protecting query images uploaded by users.
Requirement 1-a) should be always satisfied for preserving the privacy of stored images and unauthorized use.
In contrast, requirement 1-b) is needed under the assumption that cloud providers use query images on unauthorized purposes, such as collection of user’s preferences.
Thus, when users do not care about the unauthorized use of query images, the satisfaction of requirement 1-b) is not needed.

In this paper, to satisfy above all requirements, we propose a novel content-based image retrieval scheme (CBIR) for the first time.
In the proposed scheme, images are encrypted by a block-scrambling method developed for encryption-then-compression (EtC) systems\cite{EtC2,EtC3,EtC4,EtC5,EtC6}.
The encrypted images, referred to as EtC images, can be compressed with JPEG.
In addition, extended SIMPLE descriptors are combined with scalable color descriptor (SCD) to avoid the influence of image encryption.
Simulation results show that retrieval performances are the same as that of using plain images, even when the mixed use of plain images and EtC ones. 
 
\section{Related work}
\subsection{Image retrieval in encrypted domain}
Image retrieval methods in the encrypted domain are classified into two classes as shown below.
\subsection*{1) Generating descriptors from plain images}
In this class, descriptors are calculated from plain images.
For instance, SIFT-based\cite{EIR1}, SURF-based\cite{EIR4}, ORB\cite{EIR8}, MPEG-7\cite{EIR5, EIR6}, and CNN-based\cite{EIR2,EIR7,EIR9} descriptors are used in conventional schemes corresponding to this class.
To protect the content of plain images, not only plain images but also the descriptors are encrypted by a data owner.
After that, the encrypted descriptors and images are sent to a cloud server.

In this approach, data owners are required to extract descriptors and encrypt both the descriptors and plain images by themselves. 
Moreover, data owners and users have to share a common key in some schemes \cite{EIR2,EIR9}.

\subsection*{2) Generating descriptors from encrypted images}
In the second approach, descriptors are directly extracted from encrypted images by a cloud provider as well as for plain images\cite{EIR3,EIR10,EIR11,EIR12,EIR13,ESIMPLE}, where each data owner only encrypts images. 
In addition, data owners and users do not need to share keys.
The proposed retrieval scheme corresponds to this approach.

Schemes \cite{EIR3,EIR10,EIR11,EIR12,EIR13,ESIMPLE} are carried out the basis on this approach. 
Some schemes \cite{EIR11,EIR12} consider compressing images with JPEG, but their retrieval performance will be degraded due to the difference in coding parameters used for JPEG compression.
On the other hand, a scheme using extended SIMPLE descriptors extracted from EtC images does not have the limitation\cite{ESIMPLE}.
However, the mixed use of plain images and encrypted ones have never considered yet in this scheme.

\subsection{EtC image}
\label{sec:etc}
\begin{figure}[t]
\includegraphics[width=80mm]{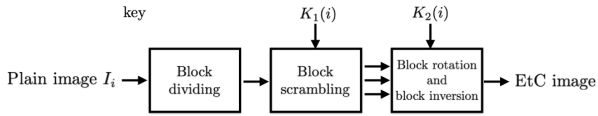}
\caption{Generation of EtC images\label{fig:enc}}
\end{figure}

\begin{figure}[t]
\begin{center}
\begin{tabular}{c}
\begin{minipage}{0.5\hsize}
  \begin{center}
   \includegraphics[width=20mm]{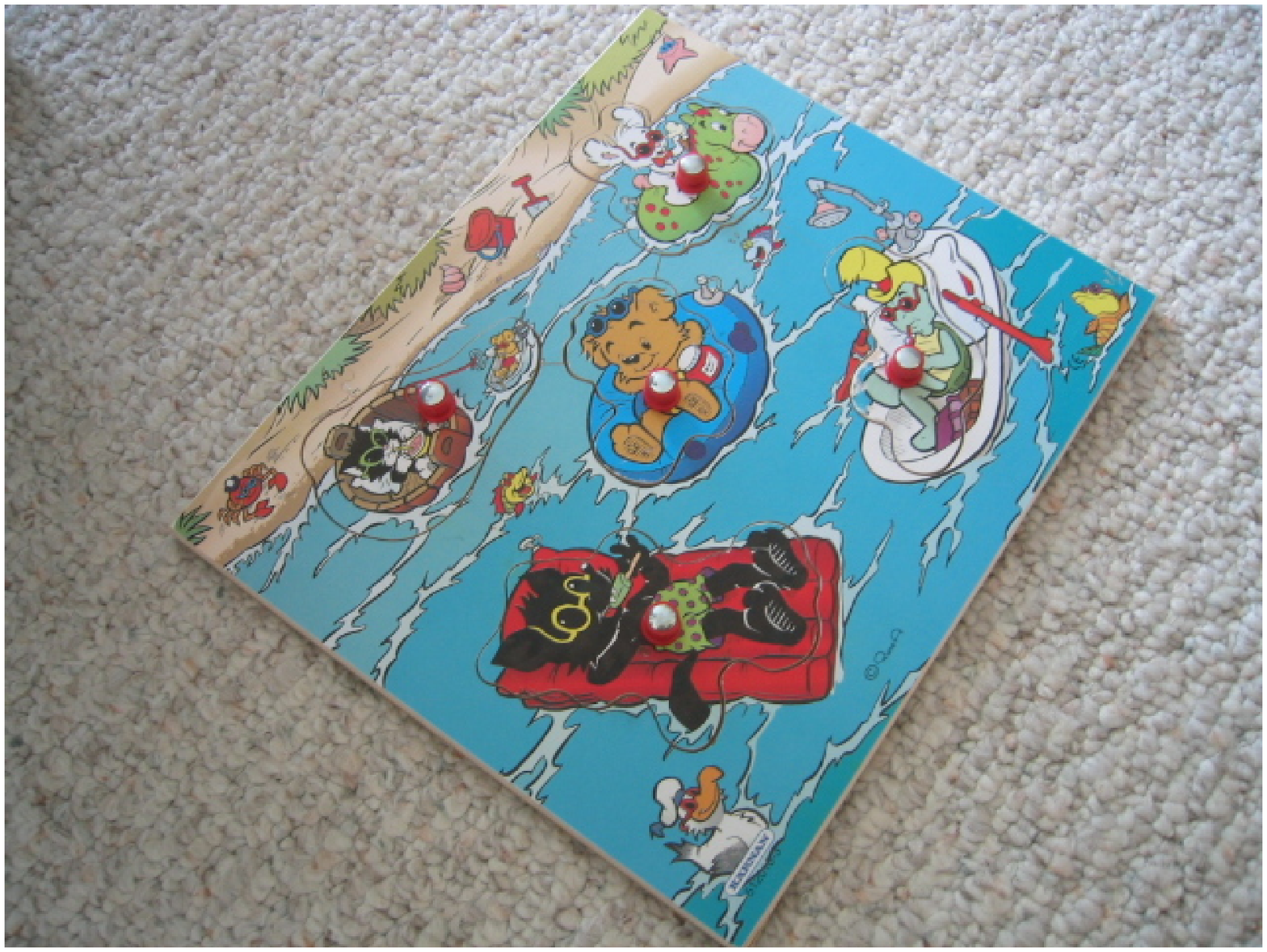}
     \hspace{20mm}(a) Plain image
  \end{center}
 \end{minipage}
\begin{minipage}{0.5\hsize}
  \begin{center}
   \includegraphics[width=20mm]{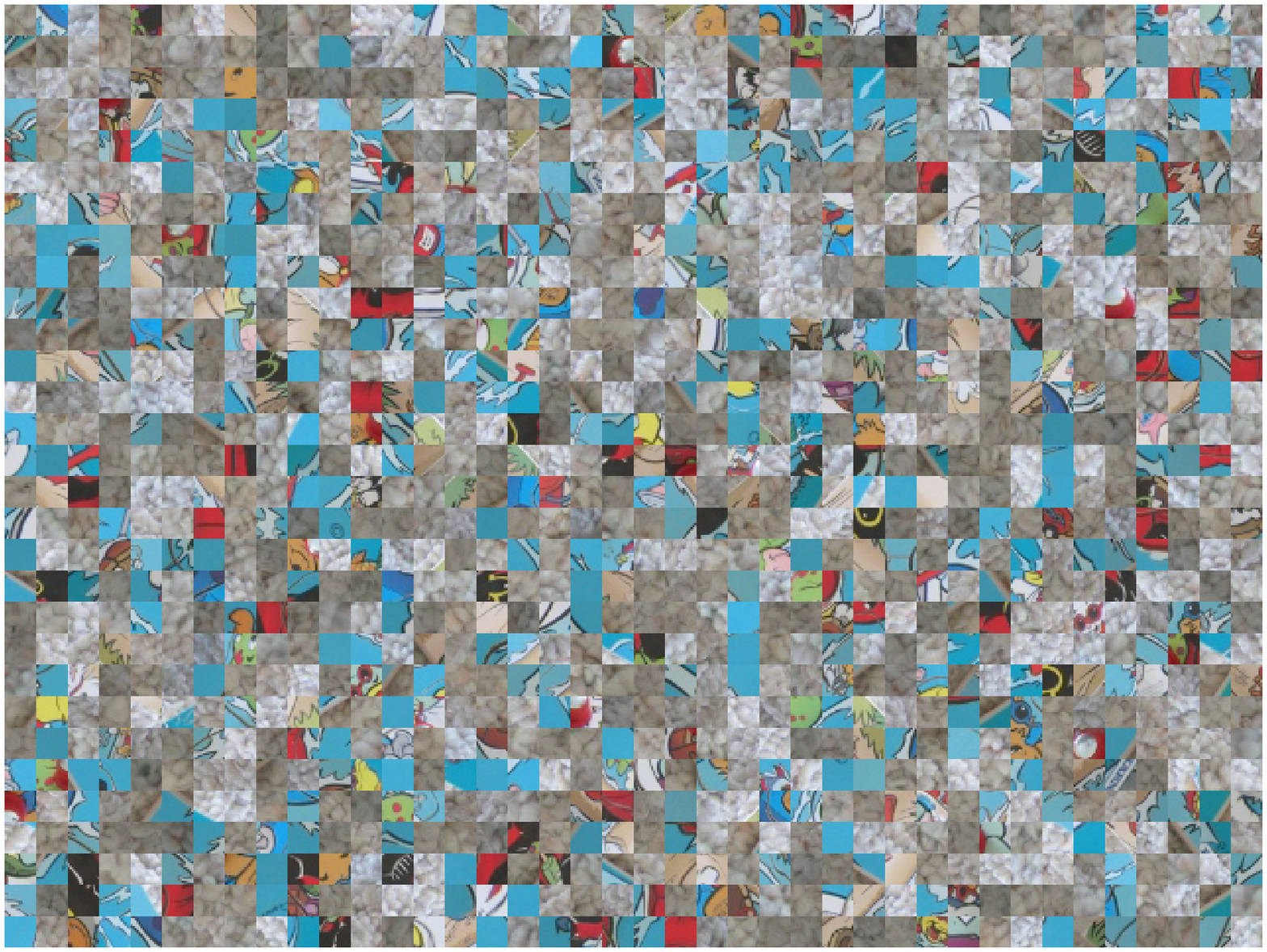}
     \hspace{20mm}(b) EtC image
  \end{center}
 \end{minipage}
\end{tabular}
\caption{Example of plain image and encrypted one\label{fig:encgen}}
 \end{center}
\end{figure}

We focus on EtC images, which have been proposed for encryption-then-compression (EtC) systems with JPEG compression \cite{EtC1,EtC2,EtC3,EtC4,EtC5,EtC6}. 
EtC images not only have almost the same compression performance as that of plain images but also enough robustness against various ciphertext-only attacks including jigsaw puzzle solver attacks\cite{EtC3,EtC4,EtC6}. 
The procedure for generating EtC images is conducted as below (see Figs. \ref{fig:enc} and \ref{fig:encgen})\cite{EtC2}.
\begin{itemize}
\item[(a)] Divide image $I_i$ with $X \times Y$ pixels into non-over-lapping $16 \times 16$ blocks. 
\item[(b)] Permute randomly $\lfloor \frac{X}{16} \rfloor \times \lfloor \frac{Y}{16} \rfloor$ divided blocks by using a random integer generated by secret key ${ K}_1(i)$.
\item[(c)] Rotate and invert randomly each divided block by using a random integer generated by secret key ${ K}_2(i)$.
\end{itemize}
In this paper,  images encrypted by using these steps are referred to as ``EtC images". 
For the retrieval of EtC images, the proposed scheme is designed to avoid the effects of two encryption operations in steps (b) and(c): block scrambling, and block rotation and block inversion.

$K_1(i)$ and $K_2(i)$ are stored as a key set $\mathbf{K}_i=[K_1(i),K_2(i)]$, which is used for the encryption of $I_i$.

\subsection{Weighted SIMPLE descriptors}
\label{sec:SIMPLE}
It is well-known that weighted SIMPLE descriptors outperform SIMPLE descriptors \cite{CBIR}.
Thus, the proposed scheme is designed on the basis of weighted SIMPLE descriptors.
The extraction weighted SIMPLE descriptors from plain images is carried out by the following steps.
\begin{itemize}
\item[a)] Extract a patch descriptor from each patch of images after selecting the positions and sizes of patches from every image. 
\item[b)] Generate a codebook with a size of $M$ from the extracted patch descriptors.
\item[c)] Calculate a histogram as a SIMPLE descriptor of each image by using the codebook and patch descriptors extracted from the image.
\item[d)] Obtain weighted SIMPLE descriptors by weighting the SIMPLE descriptors.
\end{itemize}
As mentioned above, this procedure consists of four steps: patch descriptors generation, codebook generation,  SIMPLE descriptors generation, and weighting SIMPLE descriptors.
Here, the operation of each step is summarized.
\begin{itemize}
\item Patch descriptors are extracted to generate the SIMPLE descriptor of each image and the codebook in step a).
At first, the positions and sizes of patches are determined by using a detector such as the SURF detector or random sampling.
In the case of using the SURF detector, those of each patch are determined from detected feature points and scales, respectively. 
In contrast, they are randomly selected, if random sampling is used.
After that, the type of patch descriptors extracted from selected patches is selected from MPEG-7 descriptors or MPEG-7-like descriptors\cite{CBIR}, which are represented as vectors.

\item The codebook, which is used for calculating SIMPLE descriptors, is generated from patch descriptors by using k-means clustering in step b).
The codebook consists of visual words, where  a visual word  is a vector that represents the center of each class.
Note that the size of the codebook $M$ corresponds to the number of classes in k-means clustering.

\item A SIMPLE descriptor of an image is calculated from patch descriptors of the image  and the codebook in step c).
After classifying each patch descriptor in the image into a visual word in the codebook, and the number of patch descriptors classified into the $m$th visual word corresponds to the $m$th bin in the histogram of the image, where $0 \leq m < M$. 

\item Weighting SIMPLE descriptors is achieved by the modification of each component in SIMPLE descriptors and the normalization of the modified components.
When $N$ SIMPLE descriptors are generated, the $m$th component of the $n$th weighted SIMPLE descriptor $v_n(m)$, $0 \leq m < M$, $0 \leq n < N$, is calculated as below in this paper.
\begin{equation}
v_n(m)=(1+log(tf_{(m,n)}))\times log\frac{N}{df_{(m)}},
\end{equation}
where $tf(m, n)$ represents the frequency of the $m$th visual word in the $n$th descriptor, and $df(m)$ denotes the number of SIMPLE descriptors containing the $m$th visual word in the $N$ SIMPLE descriptors.
After that, $l_2$ normalization is applied to every SIMPLE descriptor. 

\end{itemize}

%
%
\begin{figure*}[t!]
\centering
\begin{tabular}{c}
\includegraphics[width=150mm]{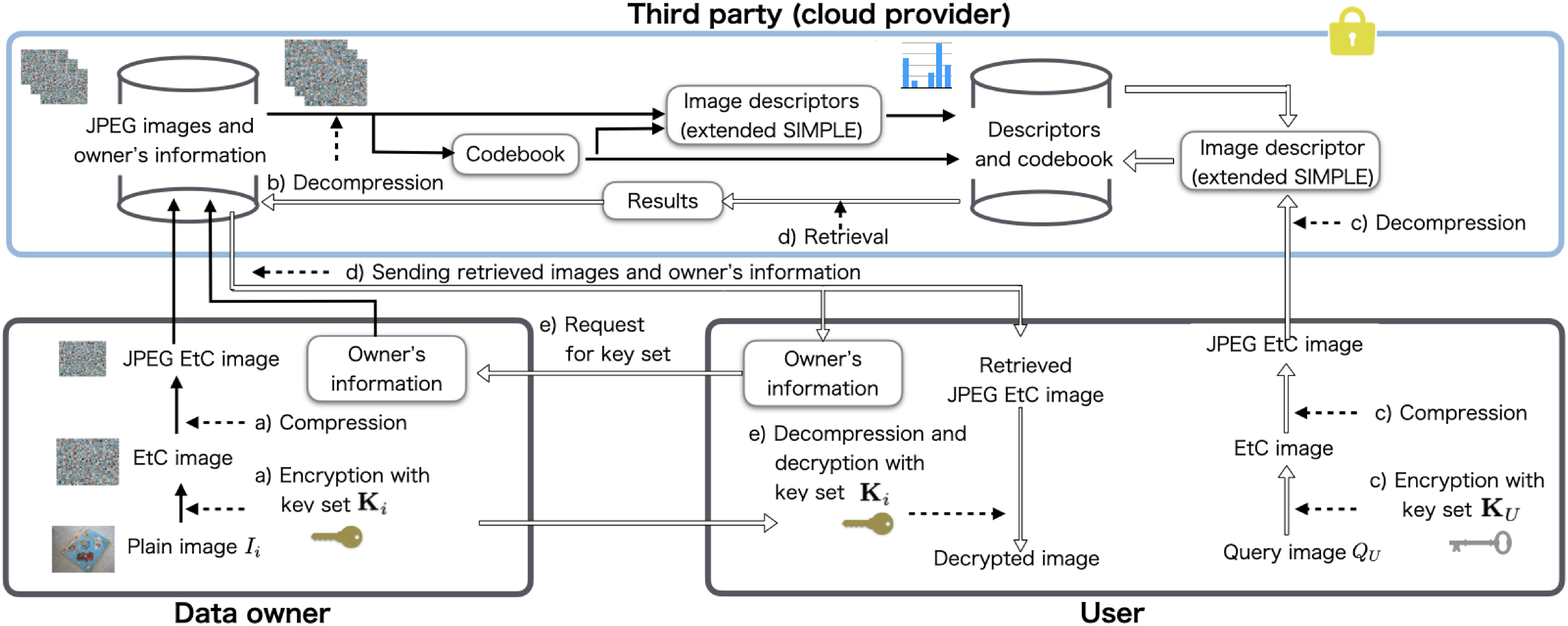}\\
(a) System model with protected query images\\
\includegraphics[width=150mm]{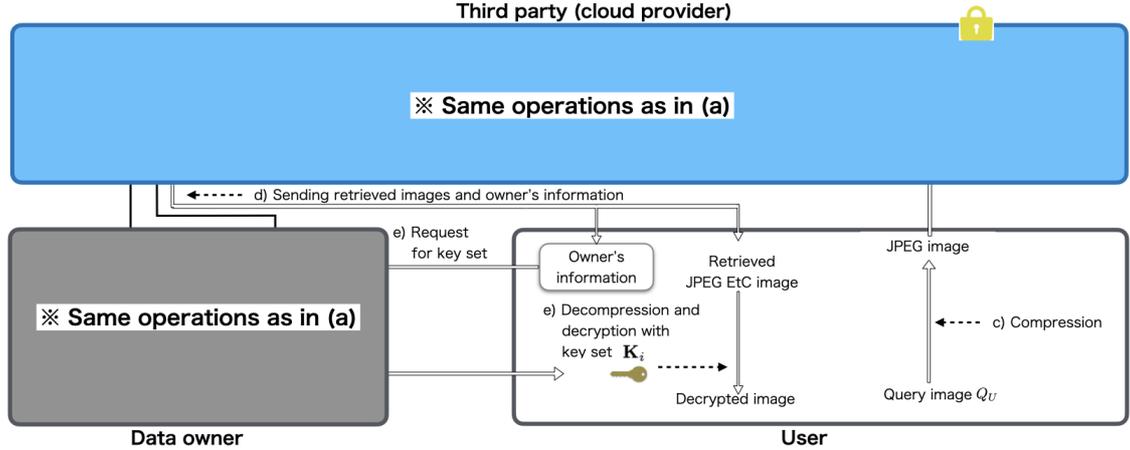}\\
(b) System model with plain query images\\
\end{tabular}
\caption{System models, where image descriptors correspond to extended SIMPLE descriptors\label{fig:scenario}}
\end{figure*}
\section{Proposed Scheme}
A novel content-based image retrieval scheme using EtC images is proposed here. 
For carrying out the image retrieval in the encrypted domain, weighted SIMPLE descriptors is modified, and SCD is used as a patch descriptor of each patch.
\subsection{System model}
Two frameworks used in the proposed scheme are shown in Figs. \ref{fig:scenario} (a) and (b).
\subsection*{1) System model with protected query images}
Each operation in the system model with protected query images, which is shown in Fig. \ref{fig:scenario} (a),  is explained as follows.
\begin{itemize}
\item[a)] A data owner generates an EtC image from plain image $I_i$ by using secret key set $\mathbf{K}_i$, and then the EtC image is compressed by the JPEG standard.
After that, the JPEG EtC image is uploaded with the owner's information to a third party.
\item[b)] The third party generates a codebook from the uploaded EtC images after decompression, and image descriptors are then calculated by using the codebook. 
Finally, the codebook and the image descriptors are stored in a database. 
\item[c)] A user encrypts query image $Q_U$ with  key set $\mathbf{K}_U$, where $\mathbf{K}_U$ can be prepared by the user.
After compressing the EtC image, the user sends the JPEG EtC image as query one to the third party.  
\item[d)] The third party calculates the image descriptor of query image, and retrieves EtC images similar to the query image in the encrypted domain. 
The retrieved images and the owner's information are returned to the user.
\item[e)] The user requests key set $\mathbf{K}_i$ from the data owner by basing on the received owner's information. 
By using key set $\mathbf{K}_i$ received from the data owner, the user decrypts the EtC images received from the third party. 
\end{itemize}
The third party not only has no visual information on images but also no secret keys in this model. 
Moreover, each image can be encrypted by using different keys.
In the other words, EtC images are able to be retrieved in this system, even when $\mathbf{K}_i \neq \mathbf{K}_U$.

\subsection*{2) System model with plain query images}
Figure \ref{fig:scenario} (b) shows a system model with plain query images.
Except for operation c), operations in this model are the same as those in the model with protected images.
Operation c) in this model is carried out as below.
\begin{itemize}
\item[c)]  A user sends query image $Q_U$ without image encryption to the third party, after applying the JPEG compression into $Q_U$.
\end{itemize}
The proposed scheme is designed by considering both the system models.

\subsection{Proposed image retrieval scheme}
To maintain the retrieval performance that using plain images can achieve, an extension of weighted SIMPLE descriptors, referred to as extended SIMPLE descriptors, is proposed to avoid the influence of block scrambling.

\subsection*{1) Extended SIMPLE descriptors}
\begin{figure}[t]
\begin{center}
\begin{tabular}{c}
\begin{minipage}{0.5\hsize}
  \begin{center}
   \includegraphics[width=25mm]{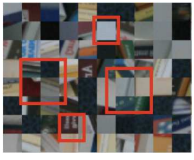}
     \hspace{20mm}(a) Random sampling
  \end{center}
 \end{minipage}
\begin{minipage}{0.5\hsize}
  \begin{center}
   \includegraphics[width=25mm]{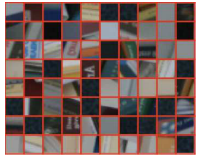}
     \hspace{20mm}(b) Extended SIMPLE
  \end{center}
 \end{minipage}
\end{tabular}
\caption{Examples of  selected patches\label{fig:patch}}
 \end{center}
\end{figure}

To generate extended SIMPLE descriptors, step a) in Sec. \ref{sec:SIMPLE} is replaced with step a’) as below.
\begin{itemize}
\item[a')] 
Divide each image into non-overlapping 16$\times$16-blocks and use each 16$\times$16-block as a patch, where 16$\times$16 corresponds to the block size of EtC images (see Fig.\ref{fig:patch} (b)).
After that, SCD is extracted as a patch descriptor from every patch. 
\end{itemize}
The use of non-overlapping 16$\times$16-blocks allows us to avoid the influence of block scrambling.
Assuming that patches are selected by random sampling and SURF detector, boundaries generated by the block scrambling operation will be included as shown in Fig.5 (a).
In contrast, every patch selected by conducting step a’) does not include these boundaries.
In addition, since extended SIMPLE descriptors are calculated by using a histogram of visual words contained in an image, block permutation in the block scrambling operation does not give any influence to the descriptors.

In addition, to be robust against block rotation and block inversion, SCD is used as a patch descriptor extracted from each 16$\times$16 blocks in step b).
SCD is represented as a vector of coefficients obtained by applying a Haar transform to a color histogram of a patch in the HSV color space.
Thus, SCD has no influence of using these operations in principle because block rotation and block inversion operations do not affect the color histogram in a patch.
In this paper, 256 coefficients are extracted from each 16$\times$16 blocks.

%
%

\subsection*{2) Retrieval process for query image} 
When the third party receives a query image from an user, the retrieval process for a query image is performed as below.
\begin{itemize}
\item[a)] Select the 16$\times$16-blocks in the query EtC image as patches after dividing the EtC image into non-overlapping 16$\times$16-blocks, and then SCD is extracted from each patch.
\item[b)] Calculate an extended SIMPLE descriptor of the query image by following steps c) and d) in Sec. \ref{sec:SIMPLE}.
\item[c)] Compute $l_2$ distance between every descriptor stored in the database and the query descriptor, and then decide similar images.
\item[d)] Send the retrieved images with owner's information to the user.
\end{itemize}
Note that these steps are the same as ones for plain images, except for obtaining patches from EtC images.

\section{Experiment}
\subsection{Experiment setup}
In this experiment,  the performance of the proposed image retrieval was evaluated by using LIRE \cite{lire}, which is an open source Java library for content-based image retrieval and supports various image descriptors.

For the comparison with the proposed scheme, five image descriptors were used: SCD\cite{scd}, color and edge directivity descriptor (CEDD)\cite{cedd}, SURF\cite{surf}, weighted SIMPLE descriptor with random sampling and weighted SIMPLE descriptor with SURF detector.
In the case of retrieval with SURF, the BOVW model and weighting term frequencies were used.

The performance was evaluated in terms of mean average precision (mAP).
To obtain mAP scores, the average of precision values was calculated for all query images.
When the number of ground truth images  is $G$, the average precision of the $q$th query
image $AP_q$ is calculated as,
\begin{equation}
AP_q=\frac{1 }{G} \sum_{n=1}^{N}\frac{TP@n}{n}\times f(n),
\end{equation}
where $N$ is the number of the images stored in the database, and $TP@n$ represents the number of the true positive matches at the rank $n$ and $f(n)=1$ if the $n$th image is a ground truth one.
Otherwise, $f(n)=0$, if the $n$th image is not.
After the calculation of average precision values for all $Q$ query images, mAP score were calculated as 
\begin{equation}
mAP=\frac{\sum_{q=0}^{Q-1} AP_q}{Q}.
\end{equation}

\begin{figure}[t]
\begin{center}
\begin{tabular}{c}
\begin{minipage}{0.23\hsize}
  \begin{center}
   \includegraphics[width=20mm]{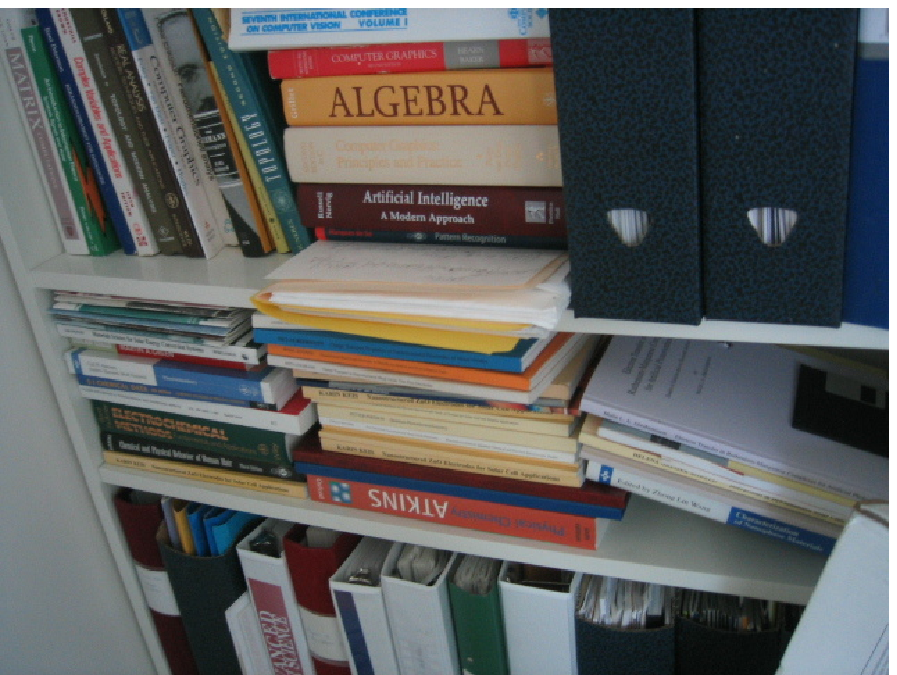}
  \end{center}
 \end{minipage}
\begin{minipage}{0.23\hsize}
  \begin{center}
   \includegraphics[width=20mm]{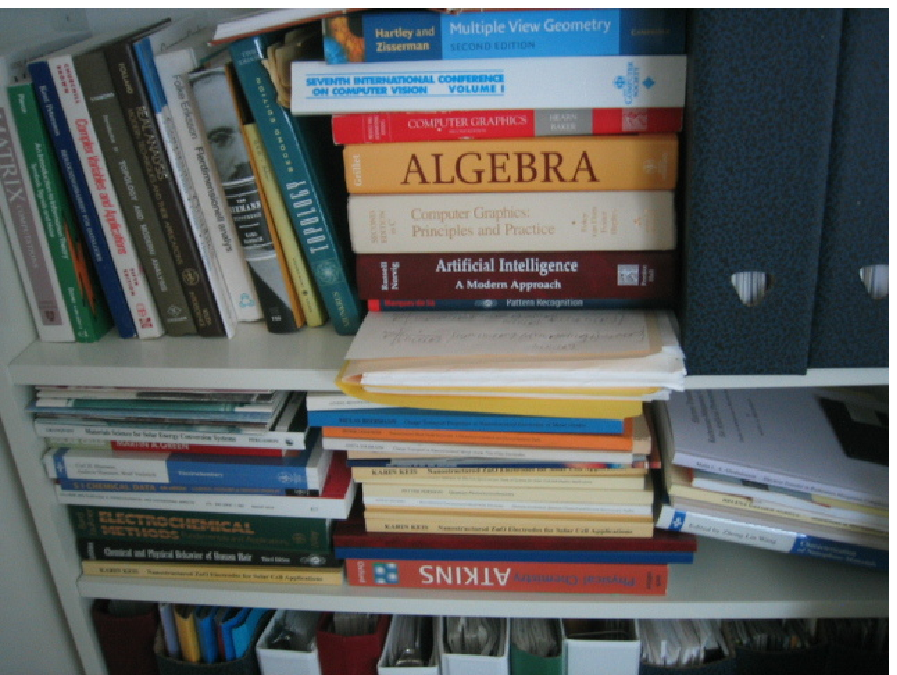}
  \end{center}
 \end{minipage}
 \begin{minipage}{0.23\hsize}
  \begin{center}
   \includegraphics[width=20mm]{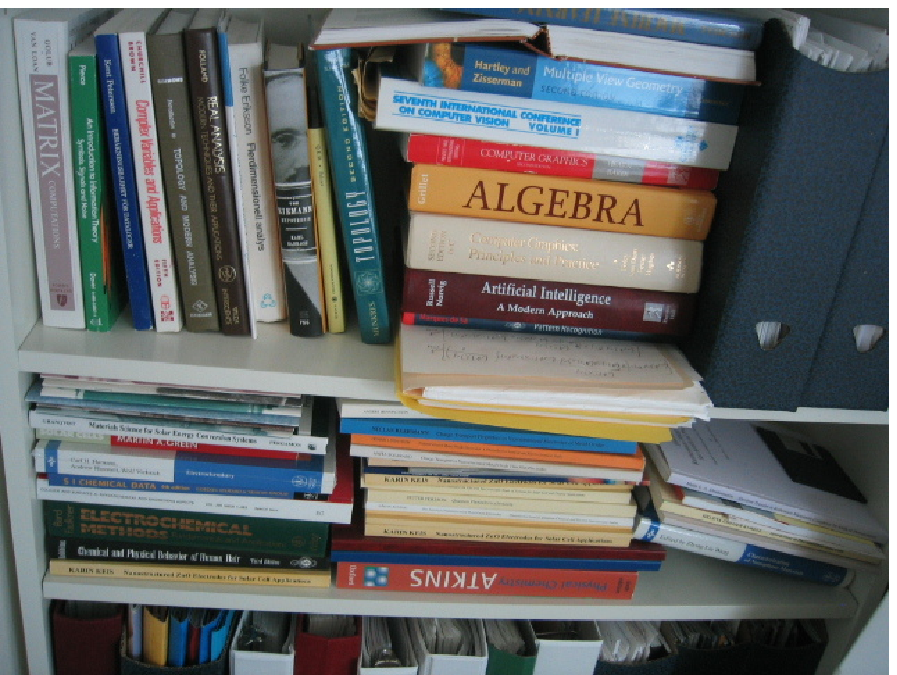}
  \end{center}
 \end{minipage}
 \begin{minipage}{0.23 \hsize}
  \begin{center}
   \includegraphics[width=20mm]{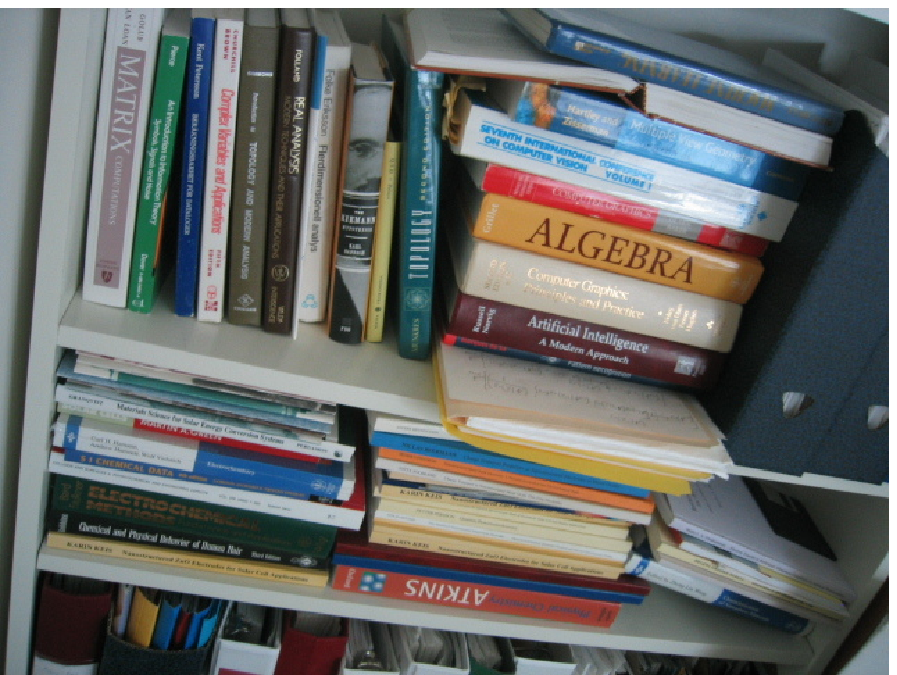}
  \end{center}
 \end{minipage}\\
 (a) Plain images \\
 
 \begin{minipage}{0.23\hsize}
  \begin{center}
   \includegraphics[width=20mm]{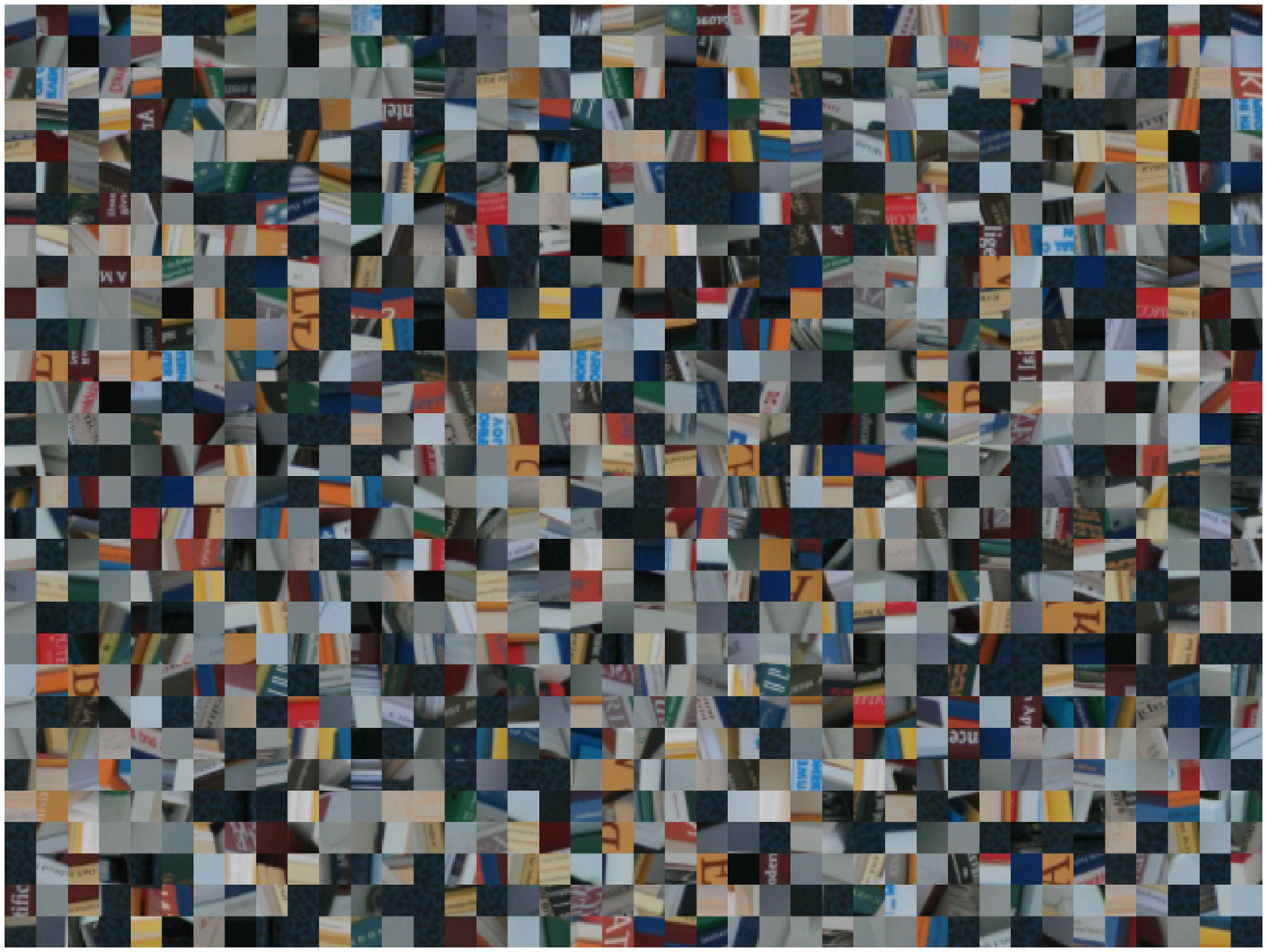}
  \end{center}
 \end{minipage}
\begin{minipage}{0.23\hsize}
  \begin{center}
   \includegraphics[width=20mm]{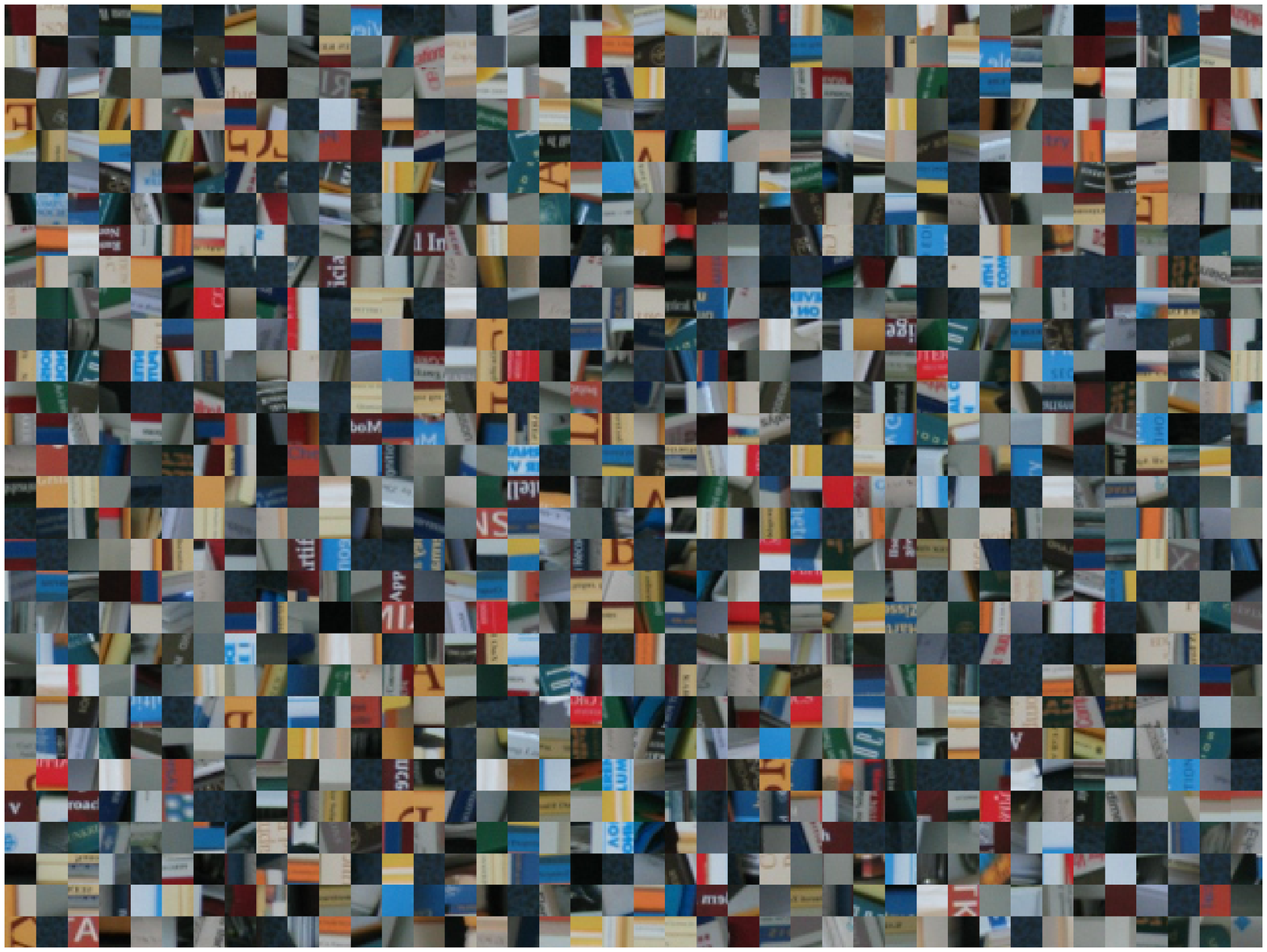}
  \end{center}
 \end{minipage}
 \begin{minipage}{0.23\hsize}
  \begin{center}
   \includegraphics[width=20mm]{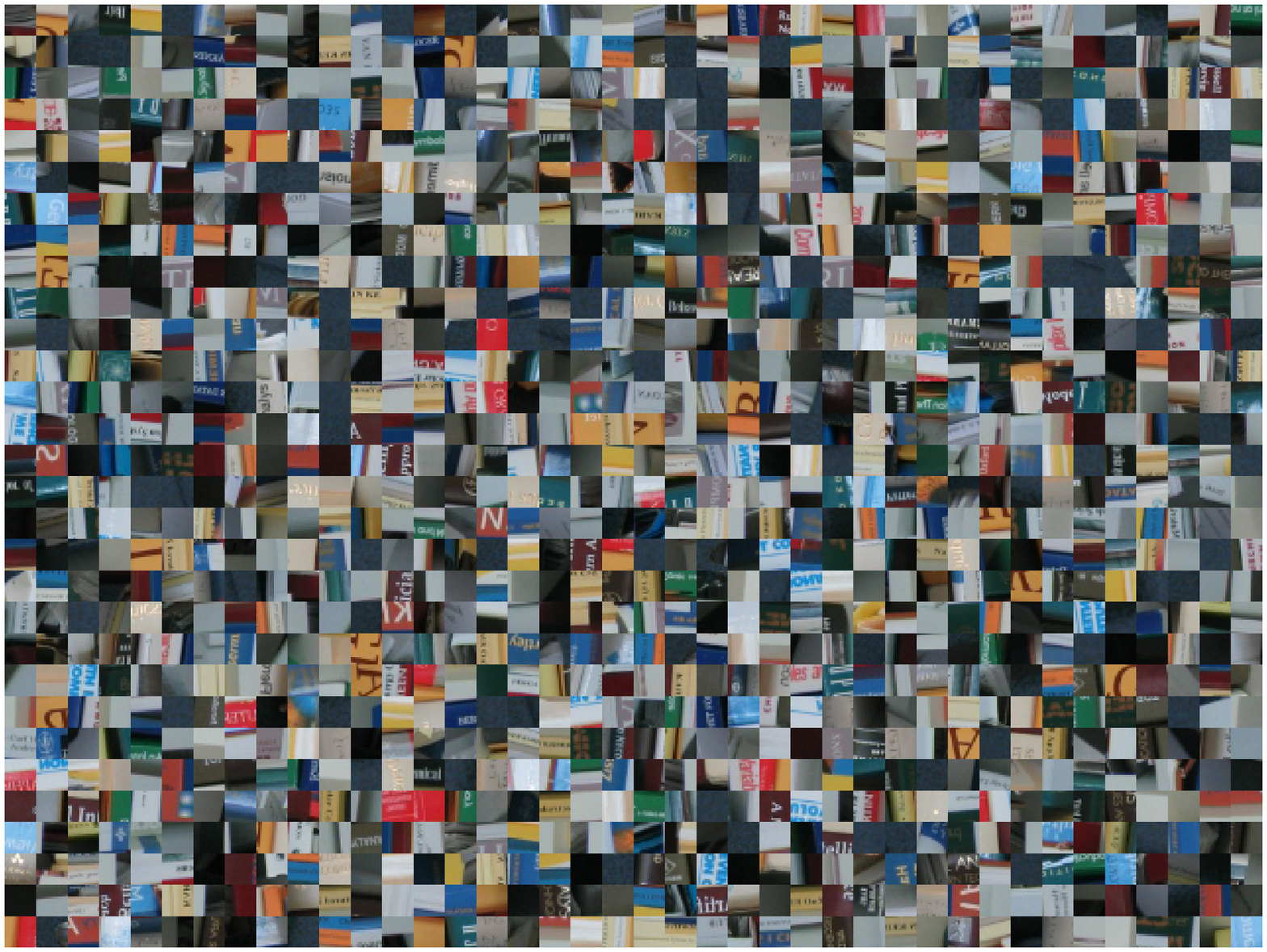}
  \end{center}
 \end{minipage}
 \begin{minipage}{0.23 \hsize}
  \begin{center}
   \includegraphics[width=20mm]{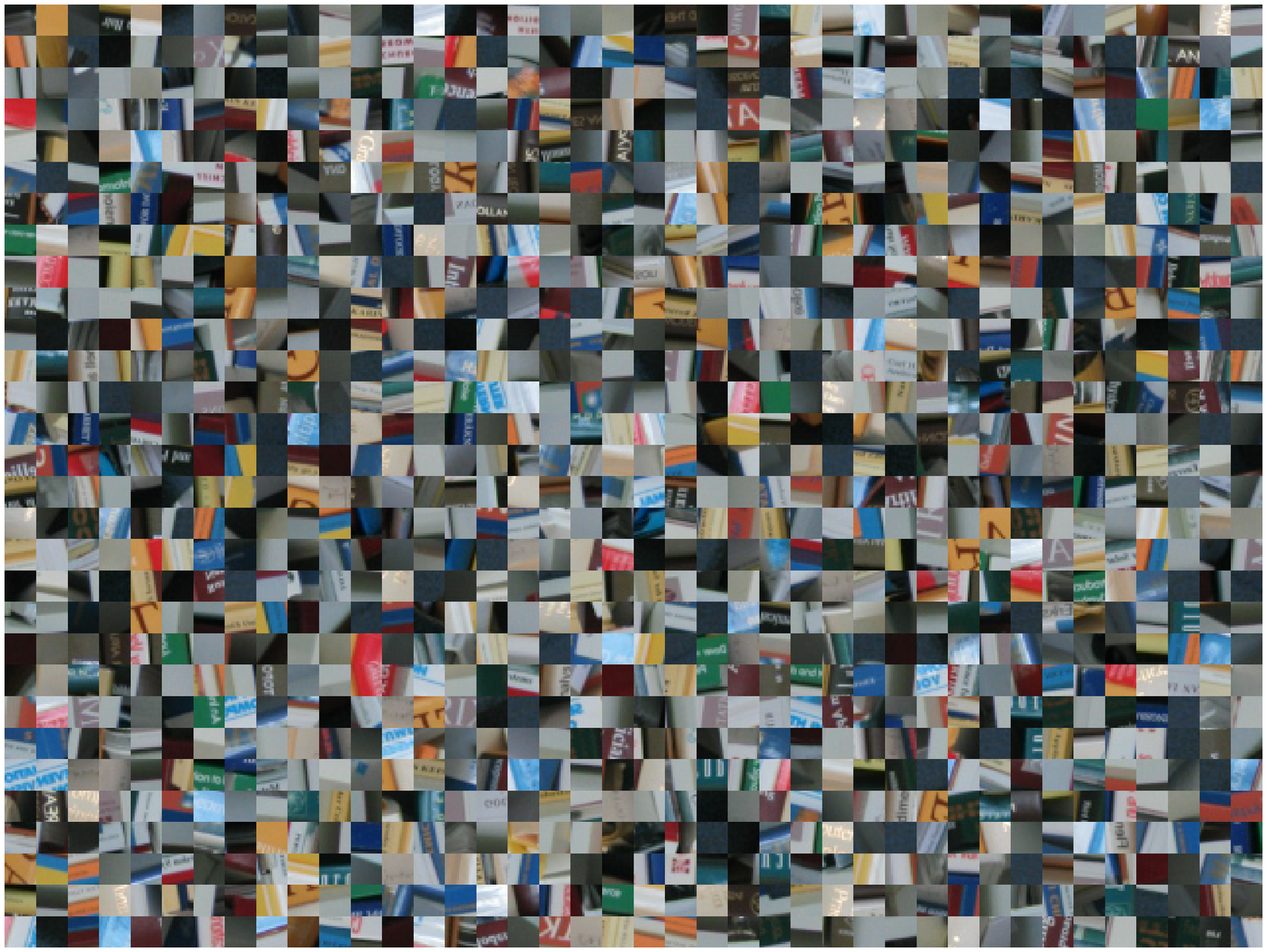}
  \end{center}
 \end{minipage}\\
(b) Corresponding EtC images \\
\end{tabular}
\caption{Image examples in group (UKbench dataset) \label{fig:exuk}}
 \end{center}
\end{figure}

\begin{figure}[t]
\begin{center}
\begin{tabular}{c}

\begin{minipage}{0.3 \hsize}
  \begin{center}
   \includegraphics[width=20mm]{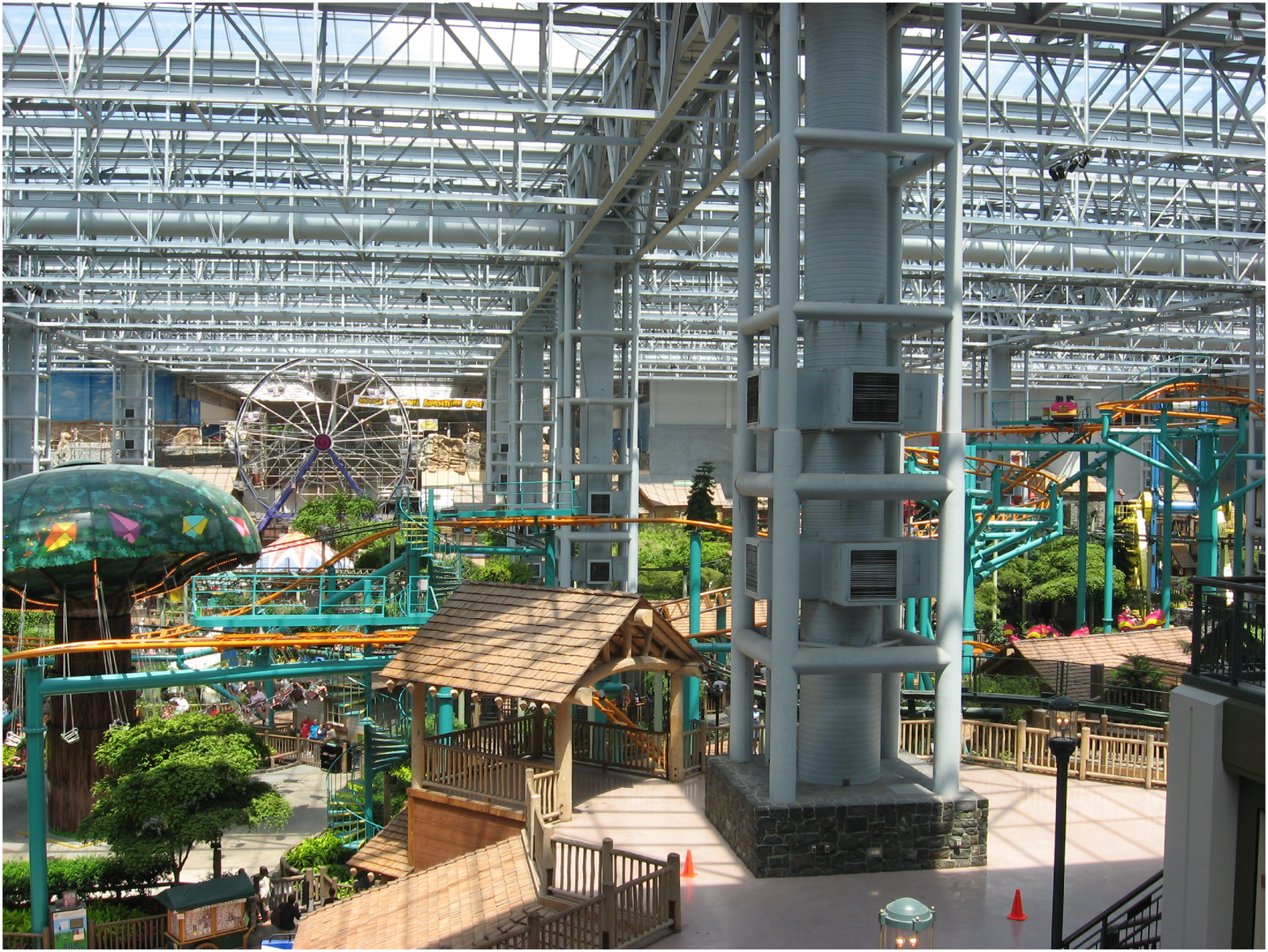}
  \end{center}
 \end{minipage}
 
\begin{minipage}{0.3 \hsize}
  \begin{center}
   \includegraphics[width=20mm]{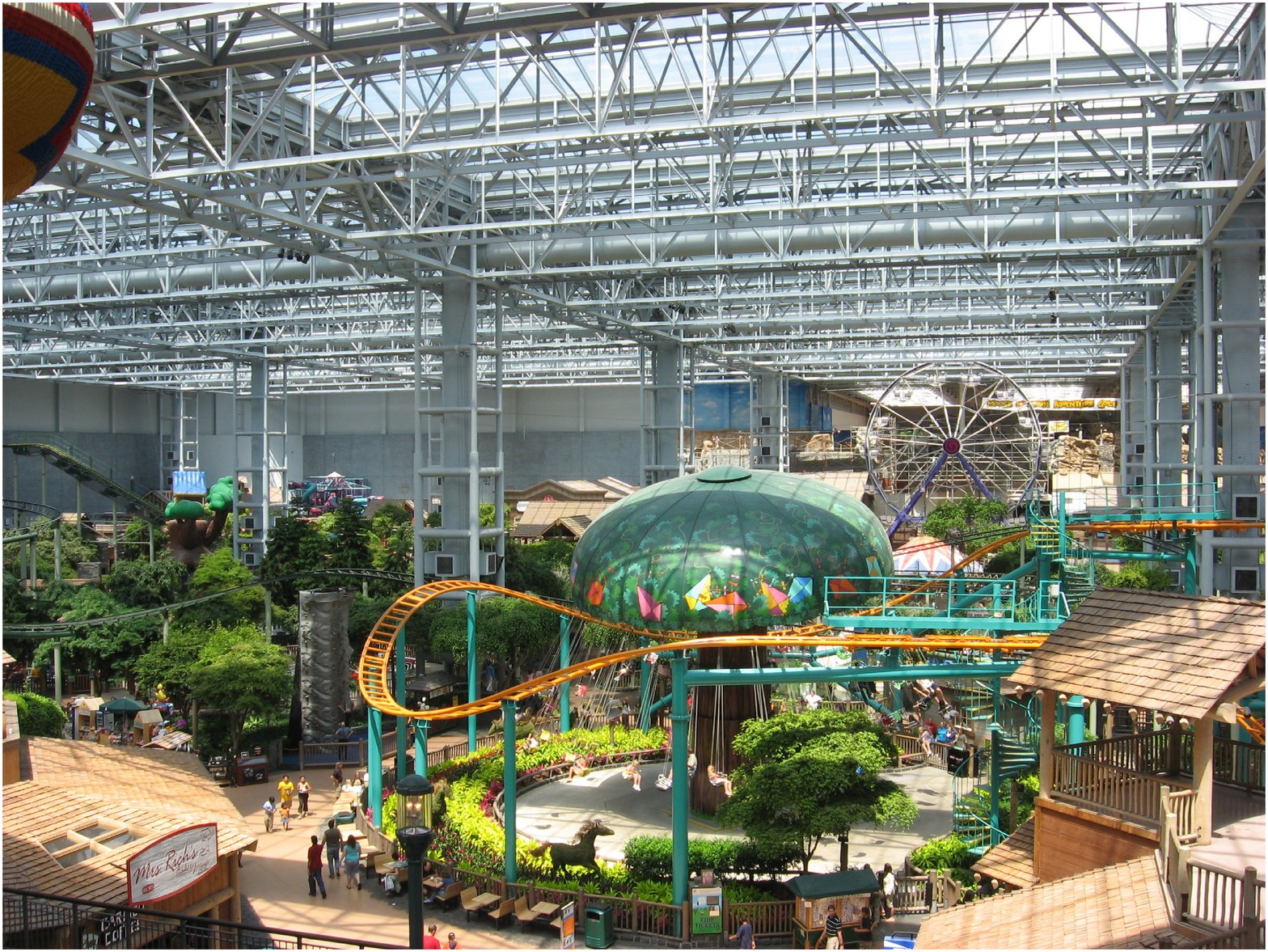}
  \end{center}
 \end{minipage}
 
 \begin{minipage}{0.3 \hsize}
  \begin{center}
   \includegraphics[width=20mm]{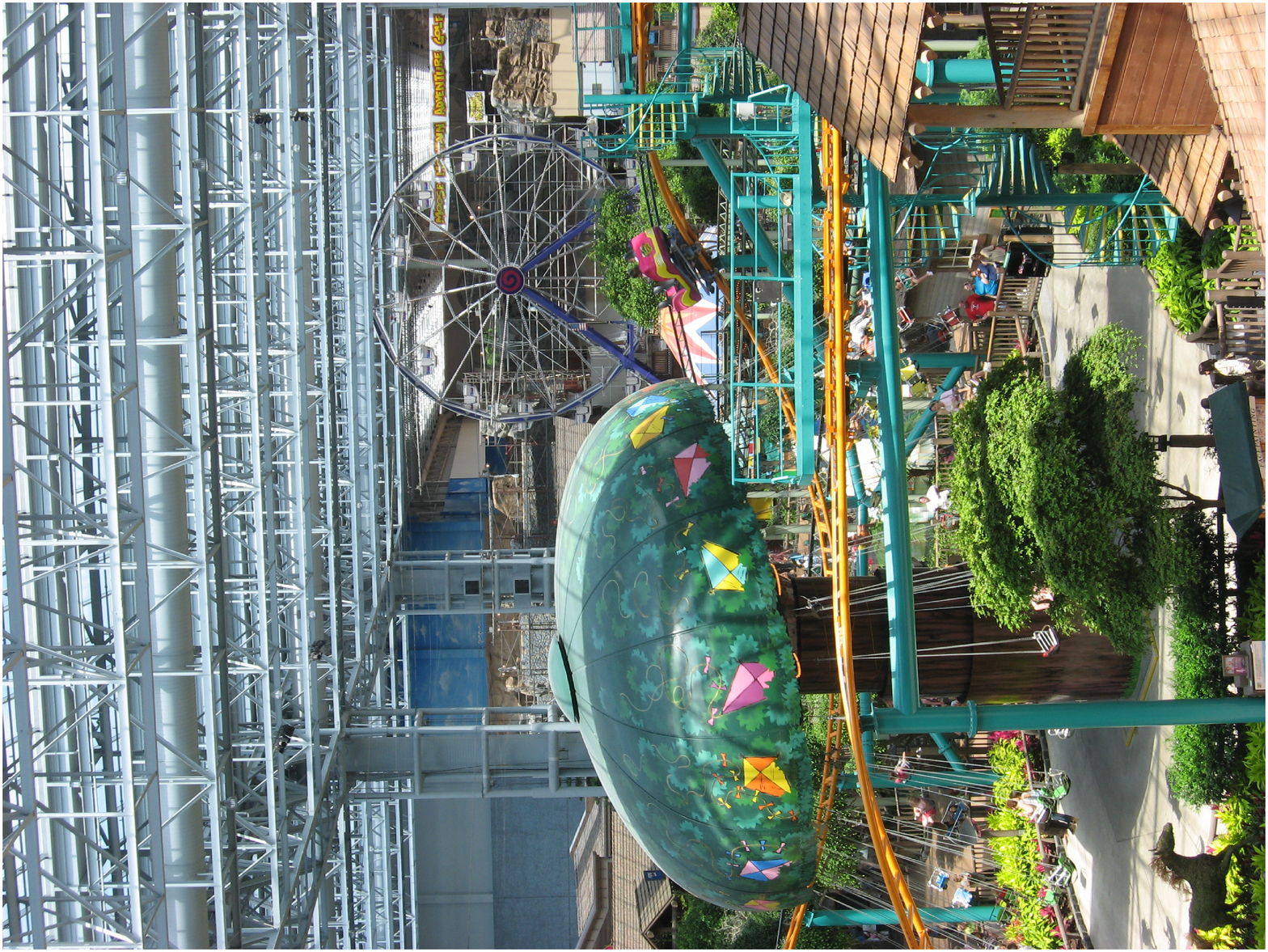}
  \end{center}
 \end{minipage}
\end{tabular} 
\caption{Image examples in group (INRIA Holidays dataset) \label{fig:exInria}}
 \end{center}
\end{figure}

In this experiment, we used two image dataset: UKbench dataset\cite{uk} and INRIA Holidays dataset\cite{INRIA} (see Figs. \ref{fig:exuk} and \ref{fig:exInria}).
UKbench dataset consists of 10,200 images with a size of 640 $\times$ 480, and the images are classified into 2,250 groups. 
Each group has four images containing a single object captured from different viewpoints and lighting conditions. 
1,000 images from No. 00000 to No. 00999 were chosen from the data set (see Fig. \ref{fig:exuk}) in this experiment. 
$N=1,000$ images were uploaded to a third party by a data owner, where the number of groups was 250, and each group consisted of 4 images, i.e., $G=4$.

In contrast, INRIA Holidays dataset contains 1,491 images, which are classified into 500 groups, and each group has at least 2 images ($2\leq G \leq 13$) (see Fig. \ref{fig:exInria}). 
In the case of using this dataset, $N=1,491$ images were stored in the third party, and $Q=500$ images were used as query ones.

\begin{table}[t!]
\caption{Relation between  types of stored and query images}
\label{tab:noteSim}
\centering
\scalebox{0.9}{
\begin{tabular}{|c|c|c|c|c|c|c|c|c|}\hline
Notation & Type of stored images & Type of query images\\\hline
``plain images vs plain images" & plain & plain \\\hline
``EtC images vs plain images" & plain & EtC \\\hline
``EtC images vs EtC images" & EtC & EtC \\\hline
\end{tabular}
}
\end{table}

\subsection{Experiment results}
\subsection*{1) Performance for UKbench dataset}
\begin{figure}[t!]
\includegraphics[width=85mm]{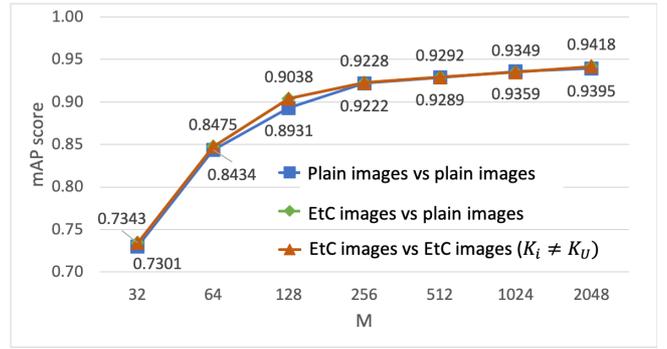}
\caption{Retrieval performance  for UKbench dataset. \label{fig:resUK}}
\end{figure}

Figure \ref{fig:resUK} shows the retrieval performances for images in UKbench dataset, where the abbreviated names shown in Tab. \ref{tab:noteSim} were used such as ``EtC images vs plain images". For instance, ``EtC images vs plain images" indicates that EtC images were stored in the database and plain images were used as query ones. 

Regarding retrieval using plain images as query images, the performance was almost the same as that under ``EtC images vs EtC images".
In addition, the retrieval performances under ``EtC images vs EtC images" had the same mAP scores as ones under ``EtC images vs plain images", even when different secret keys were used.
Moreover, Table \ref{tab:resUKorg} shows that the performances of the proposed scheme were compared with those of conventional CBIR schemes with plain images.
It was confirmed that the proposed scheme had a higher retrieval performance than the typical conventional CBIR schemes with plain images.

\subsection*{2) Performance for INRIA Holidays dataset}
\begin{figure}[t!]
\includegraphics[width=85mm]{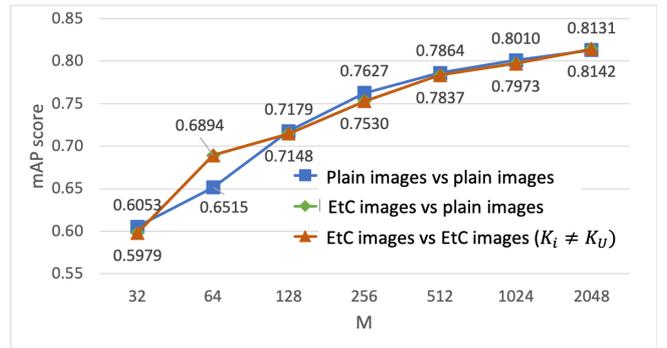}
\caption{Retrieval performance  for INRIA Holidays dataset. \label{fig:resInria}}
\end{figure}

The retrieval performances for the INRIA Holidays dataset were also evaluated, as shown in Fig. \ref{fig:resInria}.
As well as for the UKbench dataset, the performances under ``EtC images vs EtC images" had the same trend as ones under ``EtC images vs plain images".
It was also confirmed from Table \ref{tab:resInria2} that the proposed scheme had a higher retrieval performance than the conventional CBIR schemes with plain images as well.

Therefore, the proposed scheme is effective for privacy-preserving CBIR, and it allows the mixed use of encrypted and plain images as query images.
We also confirmed that the performance for compressing EtC images with JPEG had the same trend as one for compressing plain images with JPEG.

\begin{table}[t!]
\caption{Comparison with conventional CBIR methods using plain images (UKbench dataset)}
\label{tab:resUKorg}
\centering
\scalebox{1}{
\begin{tabular}{|c|c|c|c|c|c|c|c|c|}\hline
\multicolumn{2}{|c|}{Descriptor}&$M=$&mAP score\\\hline
\multicolumn{2}{|c|}{SCD\cite{scd} (plain)}& - & 0.9179\\\hline
\multicolumn{2}{|c|}{CEDD \cite{cedd} (plain)}& - & 0.8806\\\hline
\multicolumn{2}{|c|}{SURF\cite{surf}}&256&0.8304\\\cline{3-4}
\multicolumn{2}{|c|}{(plain)}&512&0.8355\\\hline
\multicolumn{2}{|c|}{Weighted SIMPLE with random sampling}&256&0.9110\\\cline{3-4}
\multicolumn{2}{|c|}{(plain)}&512&0.9262\\\hline
\multicolumn{2}{|c|}{Weighted SIMPLE with SURF detector}&256&0.8949\\\cline{3-4}
\multicolumn{2}{|c|}{(plain)}&512&0.9109\\\hline\hline
\multirow{2}{*}{Proposed}& Extended SIMPLE&256&0.9228\\\cline{3-4}
& (encrypted, $\mathbf{K}_i\neq \mathbf{K}_U$) &512&0.9292\\\hline

\end{tabular}
}
\end{table}

\begin{table}[t!]
\caption{Comparison with conventional CBIR methods (INRIA Holidays dataset)}
\label{tab:resInria2}
\centering
\scalebox{1}{
\begin{tabular}{|c|c|c|c|c|c|c|c|c|}\hline
\multicolumn{2}{|c|}{Descriptor}&$M=$&mAP score\\\hline
\multicolumn{2}{|c|}{SCD\cite{scd} (plain)}& - & 0.7524\\\hline
\multicolumn{2}{|c|}{CEDD \cite{cedd} (plain)}& - & 0.7247\\\hline
\multicolumn{2}{|c|}{SURF\cite{surf}}&256&0.6858\\\cline{3-4}
\multicolumn{2}{|c|}{(plain)}&512&0.6986\\\hline
\multicolumn{2}{|c|}{Weighted SIMPLE with random sampling}&256&0.7716\\\cline{3-4}
\multicolumn{2}{|c|}{(plain)}&512&0.7918\\\hline
\multicolumn{2}{|c|}{Weighted SIMPLE with SURF detector}&256&0.7156\\\cline{3-4}
\multicolumn{2}{|c|}{(plain)}&512&0.7338\\\hline\hline
\multirow{2}{*}{Proposed}& Extended SIMPLE&256&0.7530\\\cline{3-4}
& (encrypted, $\mathbf{K}_i\neq \mathbf{K}_U$) &512&0.7837\\\hline
\end{tabular}
}
\end{table}

\section{Conclusion}
\label{sec:conclusion}
A novel content-based image retrieval scheme allowing the mixed use of plain images and encrypted images was proposed.
To achieve a high retrieval performance, the proposed scheme was designed on the basis of weighted SIMPLE descriptors.
In addition, to avoid the influence of encrypted compressible images, called EtC images, weighted SIMPLE descriptors was also extended as extended SIMPLE descriptors under the use of SCD. Experiment results showed that the proposed scheme enables us not only to avoid the influence of the image encryption, but also to outperform conventional CBIR schemes with plain images.

\bibliographystyle{IEEEbib}
\bibliography{ref}

\end{document}